\documentclass[ preprint,11pt, amsmath,amssymb,nofootinbib]{revtex4}
\usepackage{graphicx}
\usepackage{amsmath, amssymb, graphics, epsfig}
\usepackage{dcolumn}
\usepackage{bm}
\usepackage{hyperref}
\usepackage{epstopdf}

\newcommand{\mpl}{m_{\mathrm{p}}}
\newcommand{\nc}{\newcommand}
\nc{\ba}{\begin{eqnarray}}
\nc{\ea}{\end{eqnarray}}
\newcommand\be{\begin{equation}}
\nc{\K}{{\bf k }}
\nc{\x}{{\bf x }}

\preprint{IPM/P-2010/044}

\begin{document}

\title{Curvature Perturbations and non-Gaussianities  from\\
Waterfall Phase Transition during Inflation}
\author{Ali Akbar Abolhasani$^{1,2}$}
\email{abolhasani(AT)ipm.ir}
\author{Hassan Firouzjahi$^{2}$}
\email{firouz(AT)ipm.ir}
\author{Mohammad Hossein Namjoo$^{2}$}
\email{mh.namjoo(AT)ipm.ir}
\affiliation{$^{1}$
Department of Physics, Sharif University of Technology, Tehran, Iran}
\affiliation{$^{2}$ School of Physics, Institute for Research in Fundamental Sciences (IPM),
P. O. Box 19395-5531,
Tehran, Iran}
\date{\today}

\begin{abstract}
We consider a variant of hybrid inflation where the waterfall phase transition happens during inflation. By adjusting the parameters associated with the mass of the waterfall field we arrange that the phase transition is not sharp so inflation can proceed for about 50-60 e-folds after the waterfall phase transition. We show that one can work in the limit where the quantum back-reactions are subdominant compared to the classical back-reactions. 
It is shown that significant amount of large scale curvature perturbations are induced from the entropy perturbations. The curvature perturbations spectral index is either blue or red  depending on whether the mode of interest leaves the horizon before the phase transition or after the phase transition. This can have interesting observational consequences on CMB. The non-Gaussianity parameter $f_{NL}$ is calculated to be $\lesssim 1$ but much bigger than the slow-roll parameters.

\end{abstract}
\maketitle
\section{Introduction  }
\label{intro}

Recent observations \cite{Komatsu:2010fb}
strongly support inflation as a correct theory of early universe and structure formation  \cite{Guth:1980zm}. Thanks to the precision data different inflationary models can be distinguished based on their predictions for the curvature perturbation spectral index, the amplitude of the primordial gravitational waves and the level of non-Gaussianity. 

It proved very difficult to obtain an appreciable amount of non-Gaussianity in simple models of inflation. One has to have either multiple field inflationary scenarios or non-trivial sound speed, for a review see \cite{Chen:2006nt}, \cite{Bartolo:2010qu}, \cite{Wands:2010af},  \cite{Koyama:2010xj},   \cite{Byrnes:2010em},   \cite{Suyama:2010uj}, \cite{Komatsu:2010hc} and the references therein. 
Specifically, in models of multiple field inflation when the slow-roll conditions are violated temporarily on the field space,  one may naively expect that an appreciable amount of non-Gaussianities can be produced. Careful examinations in the context of double inflation  indicate  that this may not be the case \cite{Tanaka:2010km}, \cite{Takamizu:2010xy}.
Therefore, it would be  interesting to extend these analysis to similar models where non-trivial dynamics such as a sudden change in sound speed during inflation, fields annihilations \cite{Battefeld:2008py, Battefeld:2010rf}, particle creations \cite{Barnaby:2009mc, Barnaby:2009dd, Barnaby:2010ke, Barnaby:2010sq} and phase transition \cite{Gong:2008ni, Nambu:1989eu, Li:2009sp}
happening either during inflation or at the end of inflation \cite{Sasaki:2008uc, Byrnes:2008zy, Huang:2009vk, Alabidi:2006hg, Alabidi:2006wa, Yokoyama:2008xw, Cogollo:2008bi}.

In this work we consider a variant of hybrid inflation \cite{Linde:1993cn}, \cite{Copeland:1994vg} where the waterfall phase transition happens during early stage of inflation.  By tuning the effective mass of the waterfall field, we arrange that the waterfall phase transition is mild enough such that inflation continues for a long period, say  55 e-foldings 
\cite{GarciaBellido:1996qt, Clesse:2010iz}.
To bring the effects of phase transition into cosmic microwave background (CMB) observational window, we shall assume that the phase transition happens around first few e-foldings, say first five e-folds. We follow the dynamics of both fields so our treatment is a two-field inflationary mechanism throughout. We would like to see, first, whether the entropy perturbations can induce   significant amounts of large scale curvature perturbations, and second, whether a significant amount of non-Gaussianity can be produced.  These questions \cite{Taruya:1997iv, Barnaby:2006km} attracted new interest in the literature for the model of standard hybrid inflation where the waterfall phase transition happens very efficiently at the end of inflation. As demonstrated in \cite{Lyth:2010ch}, \cite{Abolhasani:2010kr}, \cite{Fonseca:2010nk} and \cite{Gong:2010zf}
the quantum back-reactions from very small scales inhomogeneities produced during the waterfall phase transition uplift the tachyonic instability and shuts off inflation very efficiently. As a consequence, the large scale curvature perturbations are exponentially suppressed. By the same reasoning, it seems natural to ask whether this conclusion can be averted if one relaxes 
the model parameters such that  the waterfall phase transition, happening during the early stage of inflation, is not very sharp.   This is one of our main goal in this work which we shall elaborate in details in the subsequent sections. 

As a remark, somewhat related to our work here,  there have been many works in the literature concerning curvature perturbations and obtaining non-Gaussianities in models where there are local feature during inflation. In \cite{Adams:2001vc,Chen:2006xjb} 
this was translated into a sudden violation of slow roll condition in a single field model.
In \cite{Joy:2007na, Hotchkiss:2009pj} the starting model is a multiple field scenario,
where the power spectrum features are induced from slow-roll violation or phase transition due to other non-inflationary fields. However, the analysis of curvature perturbations and non-Gaussianities in these models are as  in single field inflationary models. 

The rest of the paper is organized as follows. In section \ref{hybrid} we present our background
and the classical fields equations. In section \ref{quantum} we study the quantum excitations of  the waterfall field in details and calculate the power spectrum of the entropy perturbations. In section \ref{back-reactions} we compare the quantum back-reactions and the classical back-reactions and specify the limit where the former can be safely ignored compared to the latter one. In section \ref{curvature} we study the curvature perturbations induced from the entropy perturbations in details. In section \ref{deltaN} we use the complementary $\delta N$ formalism to calculate the level of non-Gaussianity in our model. Conclusion and discussions are given in section 
\ref{conclusion} followed by an appendix describing limiting behaviors of the Bessel functions used extensively in our analysis.

\section{Waterfall Phase Transition During Inflation }
\label{hybrid}

As explained above we consider a variant of hybrid inflation model where the waterfall phase transition takes place during inflation. The potential is 
\begin{equation}
\label{pot}
V(\phi,\psi) = \dfrac{\lambda}{4} \left( \psi^2 - \dfrac{M^2}{\lambda}\right)^2 + \dfrac{1}{2} m^2 \phi^2 + \dfrac{1}{2}g^2 \phi^2 \psi^2 \, ,
\end{equation}
where $\phi$ is the conventional inflaton field, $\psi$ is the waterfall field and $\lambda$ and $g$ are two dimensionless couplings. The system has a global minimum given by $\phi=0$ and $\psi= M/\sqrt{\lambda}$. Our assumption is that the first period of inflation takes place during $\phi_c < \phi < \phi_i$ where $\phi_i$ is the initial value of the inflaton field and $\phi_c = M/g$ is the critical value of $\phi$ where the waterfall field becomes instantaneously massless. 
Unlike conventional hybrid inflation, we assume that the waterfall field is not heavy so it can also slowly roll down during the first stage of inflation. 
For $\phi<\phi_c$ the waterfall becomes tachyonic triggering an instability in the system. Again, unlike standard hybrid inflation, we assume that this phase transition is mild enough such that it will take a long time for the waterfall field to settle down to its global minimum. This provides the second stage of inflation. Our parameters should be such that the second stage of inflation is long enough say 55 e-foldings or so. 

It is well-known that the potential (\ref{pot}) with the discrete $Z_2$ symmetry is plagued with the domain-wall formations at the end of inflation \cite{Felder:2000hj}
which are dangerous cosmologically.
To get rid of this problem one has to consider $\psi$ to be a complex scalar field so at the end of inflation
cosmic strings are produced which can be safe cosmologically. However, the process of topological defect formation is beyond the scope of this work and we shall proceed with $\psi$
to be real as is customary in many models of hybrid inflation in the literature. 

\subsection{The Background Fields Dynamics}
\label{background}
Here we study the classical evolutions of background fields $\phi$ and $\psi$ before, during and after the phase transition.  The background space-time metric, as usual, is
\ba
ds^2 = - dt^2 + a(t)^2 d\x^2 \, ,
\ea
where $a(t)$ is the scale factor.

It is more convenient to use the number of e-foldings, $N$,  as the clock $d N = H \, d t$. The background fields equations are 
written as 
\ba
\label{phi-c}
\phi'' + (3-\epsilon)  \phi' + \left(\alpha +g^2  \dfrac{\psi^2 }{H^2} \right) \phi =0 \\
\label{psi-c}
\psi'' + (3-\epsilon)  \psi' + \left(-\beta +g^2 \dfrac{\phi^2}{H^2}+\lambda \dfrac{\psi^2} {H^2}\right) \psi =0 \, ,
\ea
where the prime denotes the differentiation with respect to the number of e-foldings and 
$\epsilon\equiv  -\dot H/H^2$ is the slow-roll parameter which is assumed to be very small by construction. Also the dimensionless parameters $\alpha$ and $\beta$ are defined as
\ba
\label{alpha}
\alpha = \dfrac{m^2}{H^2} \qquad , \qquad \alpha_0= \frac{m^2}{H_0^2}=
\dfrac{3 \lambda m^2 \mpl^2}{2 \pi M^4} 
\ea
and
\ba
\label{beta}
\beta = \dfrac{M^2}{H^2} \qquad , \qquad   \beta_0 \equiv \dfrac{M^2}{H_0^2}=\dfrac{3 \lambda  \mpl^2}{2 \pi M^2}\, ,
\ea
where $\mpl= 1/G$ for $G$ being the Newton constant and $H_0 \equiv \sqrt{2 \pi/3 \lambda} \,  M^2 / \mpl$. Here the quantities $\alpha_0$ and $\beta_0$ respectively  
represent the values of the parameters $\alpha$ and $\beta$ in the limit when we neglect the variation of $H$ during inflation so  $H\simeq  H_0$, $\alpha \simeq \alpha_0$ and $\beta \simeq \beta_0$. However, in order  to get accurate enough solutions
it is important to consider the running of $H$ in our analytical treatments. We work in the limit where $\alpha \ll 1$ corresponding to a light $\phi$ field. Also since we are interested in a mild phase transition, we take $\beta \lesssim 1$. This is in contrast to standard hybrid inflation model with a sharp phase transition at the end of inflation where $\beta \simeq \beta_0 \gg 1$. 

To simplify the notation, we take the critical point as the reference point and define
 $n \equiv N -N_c$. We use the convention  that at the start of inflation for $\phi=\phi_i$, $N=0$,  at the time of phase transition $N= N_c$ and at the end of inflation $N= N_e$. With this convention $n<0$ before the phase transition whereas $n>0$ afterwards. 
\\
Let us  for the  moment assume that $\alpha $ and $\beta$ are constants. At the end of analysis, we will include the running of $\alpha$ and $\beta$ effectively in the analysis. 
Also we are in the limit where $g^2 \psi^2/H^2 \ll \alpha$, i.e. the back-reaction of the waterfall field on the inflaton field is small during inflation. With these assumptions one can  easily solve Eq. (\ref{phi-c})  to get
\ba
\label{phif}
\phi(n) \simeq \phi_c \exp \left(-r_0 ~n \right)  
\ea
with
\ba
r_0 = \left( \dfrac{3}{2}- \sqrt{\dfrac{9}{4}-\alpha_0}\right) \simeq \frac{\alpha_0}{3} 
\ea
Equivalently, one also has
\ba
\label{nc}
N _c \simeq \dfrac{1}{r_0} \ln \left( \dfrac{\phi_i}{\phi_c}\right) \, .
\ea

As explained before, one can not neglect the running of $\alpha$ and $\beta$ which results in significant errors in the results above.  Here we take into account the running of $\alpha$ and 
$\beta$ . For this purpose, we consider the next leading term in $H$ from the  Friedmann equation  
\ba
\label{runningH}
H^2 \simeq \dfrac{8 \pi}{3 \mpl^2} \left(\dfrac{M^4}{4 \lambda} +\dfrac{1}{2} m^2 \phi^2 \right) \, .
\ea
It turns out that the correction to $\alpha$ are less crucial as compared to $\beta$ and an overall averaging would suffice.  We define the averaged $H^2$ via
\ba
\overline{H^2} = \frac{1}{N} \int_0^N H^2(n) d n  \, .
\ea
Plugging the result  from Eq. (\ref{runningH}) and performing the integral, one obtains
\ba
\label{H-run}
\overline{H^2}(N)  \simeq H_0^2 \left( 1+ \frac{2 \pi \phi_c^2}{N \mpl^2}
\right) \, .
\ea
Using this averaged value of $H^2$ into the definition of $\alpha$  results in 
\ba
\label{alpha-ef}
{\alpha} \simeq \alpha_0 \left(1- \dfrac{2 \pi}{N_e}  \dfrac{\phi_i^2}{\mpl^2}\right) \, ,
\ea
in which $N_e$ is total number of e-foldings. This equation modifies $r_0$ to an effective value of $r= \alpha/3$.  Now plugging Eqs. \eqref{phif} and \eqref{H-run}
in the definition of $\beta$  (with $r_0\rightarrow r$) one obtains the first order correction 
to $\beta$
\ba
\label{beta-modified}
\beta \simeq \beta_0 \left( 1-\Gamma e^{-2 r \, n}\right)  \, ,
\ea  
where $\Gamma \equiv \dfrac{4 \pi}{3} \alpha_0 \left( \dfrac{\phi_c}{m_{pl}} \right)^2 \ll 1$.

Let us now turn to the dynamics of $\psi$ field. As it can be confirmed from our full numerical results we are in the limit where $\psi^2 / H_0^2 \ll \beta /\lambda$
so the equation for $\psi$ simplifies to
\ba
\label{psi-substituted}
\psi''+3\psi' +\beta \left( e^{-2r\,n} -1\right)\psi =0 \, .
\ea
Using the modified form of $\beta$ from Eq. (\ref{beta-modified}), the equation of motion for   $ \psi$ modifies to
\ba
 \psi^{''} + 3  \psi^{'}  - \beta_0 \left[1-(1+\Gamma) e^{-2 r n}+\Gamma e^{-4 r n} \right] \psi =0 \, .
\ea
The solutions of the above equation are given in terms of the  Whittaker functions
$ e^{(r-3/2) n} M_{\kappa , \mu} \left( \dfrac{\sqrt{\beta_0 \, \Gamma}}{r} e^{-2r n}\right)$
and 
$ e^{(r-3/2) n} W_{\kappa , \mu} \left( \dfrac{\sqrt{\beta_0 \, \Gamma}}{r} e^{-2r n}\right)$, 
in which 
$\kappa \equiv \dfrac{\sqrt{\beta_0}(\Gamma+1)}{4 r \sqrt{\Gamma}}  $, $\nu \equiv \dfrac{1}{r} \sqrt{\beta_0 + 9/4}$ and $\mu \equiv \nu /2 $.
In the solution above one can check that $\kappa \gg 1$ and we can approximate the Whittaker functions with the Bessel functions and 
\ba
\label{psisol1}
\psi \simeq e^{-3 n/2 } \left[ c J_{\nu} \left( \dfrac{\sqrt{\tilde{\beta}}}{r} e^{- r \, n} \right) +c^{'} Y_{\nu} \left( \dfrac{\sqrt{\tilde{\beta}}}{r} e^{- r \, n} \right)  \right] \, ,
\ea
where $\tilde{\beta} \equiv \beta_0 (1+\Gamma) $.

For our range of parameters one can easily show that $\nu >\dfrac{\sqrt{\tilde{\beta}}}{r} e^{- r \, n} \gg 1$ and in this limit $Y_\nu(x)$ is much larger than $J_\nu(x)$. After imposing the initial conditions the contributions from $J_\nu$ in Eq. (\ref{psisol1}) 
becomes negligible and the classical trajectory can be approximated by
\ba
\label{psiapp1}
\psi \simeq \psi_i ~ e^{-3N/2} ~\dfrac{Y_{\nu} \left(\frac{\sqrt{\tilde{\beta}}}{r} e^{-rn}\right)}{Y_{\nu} \left(\frac{\sqrt{\tilde{\beta}}}{r} e^{rN_c} \right)} \, .
\ea 
We have checked that this analytic formula for $\psi$ is in good agreement with the results obtained from the full numerical analysis.

\section{Dynamic of Quantum fluctuations}
\label{quantum}

In this section we study the dynamics of  waterfall fields quantum fluctuations which play the role of entropy perturbations. The goal is to calculate the curvature perturbations induced from these entropy perturbations which would be the subject of the studies in section \ref{curvature}.
As we shall see the adiabatic curvature perturbation from the inflaton field is subdominant compared to the curvature perturbation induced from the entropy field.

The equation governing the dynamics of waterfall quantum fluctuations, $\delta \psi_{\mathbf{k}}$, in momentum space $\mathbf{k}$  is
\ba
\delta \psi_{\mathbf{k}}'' + 3 \delta \psi_{\mathbf{k}}' + \left( \dfrac{k^2}{a^2 \mathrm{H}^2} -\beta +g^2 \dfrac{\phi^2}{H^2} \right) \delta\psi_{\mathbf{k}} =0 \, .
\ea
By substituting the equation of $\phi$ field from {\bf Eq.} \ref{phif} one has
\ba
\label{deltapsi-de}
\delta \psi_{\mathbf{k}}'' + 3 \delta \psi_{\mathbf{k}}' + \left( \dfrac{k^2}{a^2 \mathrm{H}^2} + \beta \left( e^{-2 r n}-1 \right) \right) \delta\psi_{\mathbf{k}} =0 \, .
\ea

Since the effective mass of $\delta \psi_{\mathbf{k}}$ is at the same order as $H$, that is 
$\beta \lesssim 1$, one can neglect the term containing $\beta$ in Eq. (\ref{deltapsi-de}) for the sub-horizon perturbations and  the solution of the $\delta \psi_\K$ excitations inside the horizon is given in terms of the Hankel functions $ H^{(1)}_{3/2} (k e^{-n}/k_c)$ and $ H^{(2)}_{3/2} (k e^{-n}/k_c)$ where  $k_c \equiv He^{N_c}$ is the critical mode which exits the horizon at the moment of phase transition.  We require  that deep inside the horizon the solutions start from the Bunch-Davis
vacuum  
\ba 
\delta \psi_k^- \to \dfrac{e^{-ik \tau} }{a \sqrt{2k}}  \qquad \mathrm{as} 
\quad -k\tau \to \infty \, ,
\ea
where $\tau$ is the conformal time, $d t = - a d \tau$. With this initial condition the incoming solution for the modes inside the horizon is obtained to be 
\ba
\label{deltapsi-}
\delta \psi_\K^-(n) =-\sqrt{\dfrac{\pi}{4 k_c}} ~  e^{-N_c}  ~  e^{-3n/2} ~ H^{(1)}_{3/2} \left( \dfrac{k}{k_c}~e^{-n} \right)  \, .
\ea
As can be seen from this expression, the amplitude of the quantum fluctuations  at the time of horizon crossing $n_*$ when $k= e^{n_*} k_c$ is given by
\ba
\left \vert \delta \psi _{\mathbf{k *}}  \right \vert \simeq \dfrac{H}{\sqrt{2k^3}} \, .
\ea
Note that here and below an asterisk represents the values of the corresponding quantities at the time of horizon crossing.

After horizon crossing one can neglect the term containing $k^2$ in Eq. (\ref{deltapsi-de})  and the equation for $\psi_\K $ becomes identical to the background $\psi$ equation, { Eq.} \eqref{psi-substituted}, with the answer similar to Eq. \ref{psiapp1}
\ba
\label{deltapsi+}
\delta \psi_k^+ \simeq e^{-3 n/2 } \left[ c_1 J_{\nu} \left( \dfrac{\sqrt{\tilde \beta }}{r} e^{- r \, n} \right) +c_2 Y_{\nu} \left( \dfrac{\sqrt{\tilde \beta }}{r} e^{- r \, n} \right)  \right] \, .
\ea
We need to fix the constants of integrations $c_1$ and $c_2$ by imposing the matching conditions connecting the outgoing solution $\delta \psi_\K^+$ to the incoming solution $\delta \psi_\K^-$. The matching condition is performed at $n=n_m$  when the term containing $\beta$ in Eq. (\ref{deltapsi-de})  becomes comparable to the term containing $k^2$ which results in the following equation for $n_m$
\ba
\label{nm}
\left(\dfrac{k}{k_c}\right)^2 e^{-2n_m}=\beta \vert e^{-2rn_m}-1 \vert \simeq 2 \beta_0 r \vert n_m \vert \, . 
\ea
From the above equation one observes that $n_m$ can be either positive or negative for the physically relevant modes. This means that  the time of matching can occur either before or after the waterfall phase transition. Furthermore, comparing Eq. (\ref{nm}) with the equation of $n_*$, that is  $e^{n_*}= k/k_c$,  one concludes that $n_m>n_*$  and for a given mode
the time of performing the matching condition is always after the time when
that mode leaves the horizon.  

To further simplify the analysis of matching conditions we can neglect the running of 
$n_m$ compared to $e^{n_m}$and  replace Eq. (\ref{nm}) with the following simpler equation 
\ba
\label{approxmatch}
\left(\dfrac{k}{k_c}\right) e^{-n_m}=\sqrt{2 \,  \beta_0 \, r} \, .
\ea 

Imposing the conditions $\delta \psi_\K^-(n_m)=\delta \psi_\K^+(n_m) $ and 
$\delta \psi_K^{'-}(n_m)=\delta \psi_K^{'+}(n_m)$  we can fix $c_1$ and $c_2$
\ba 
\label{c}
c_1=-\sqrt{\dfrac{\pi}{k_c}}~\dfrac{\pi e^{-N_c}}{4r} \left[\sqrt{\tilde \beta} e^{-r n_m}~H_{3/2}^{(1)} ~Y_\nu^{'}-\dfrac{k}{k_c} e^{-n_m} ~H_{3/2}^{'(1)}~ Y_\nu \right]
\nonumber \\
c_2=+ \sqrt{\dfrac{\pi}{k_c}}~\dfrac{\pi e^{-N_c}}{4r} \left[\sqrt{\tilde \beta}e^{-r n_m}~ H_{3/2}^{(1)} ~J_\nu^{'}-\dfrac{k}{k_c}e^{-n_m}~ H_{3/2}^{'(1)} ~J_\nu \right] \, .
\ea
Here primes  denote derivatives with respect to the argument of the Bessel and the Hankel
functions. Also the arguments of the Hankel and the Bessel functions, respectively, are the same as those in \eqref{deltapsi-} and \eqref{deltapsi+} with $n=n_m$. 
We have checked that with these values of $c_1$ and $c_2$, our analytical solutions  \eqref{deltapsi-} and \eqref{deltapsi+} are in very good agreements with the results obtained from the full numerical analysis.

In the  Appendix we presented approximate formulae for $c_1$ and $c_2$. Considering the fact that  we are in the limit where $\nu \gtrsim \sqrt{\beta_0 (\Gamma + 1)} e^{-r n}/r \gg 1$ 
and using the approximate expressions  for $c_1$ and $c_2$ one can  check that 
the term containing $J_\nu$ in \eqref{deltapsi+} becomes subdominant and
\ba 
\label{approxdeltapsi+}
\delta \psi_k^+ \simeq c_2  \, e^{-3n/2} \, Y_\nu \left( \dfrac{\sqrt{\tilde \beta}}{r} e^{-r\,  n}\right) .
\ea

Now it is time to compute the power spectrum of entropy perturbations, ${\cal S}$. 
Following the prescription of \cite{Gordon:2000hv} we can perform a local rotation from 
$\phi-\psi$ field space into $\sigma-s$ field space where $d \sigma \equiv \cos \theta d \phi + \sin \theta d \psi $ represents the adiabatic field tangential to the classical trajectory while
$ds \equiv \cos \theta d \psi - \sin \theta d \phi$ represents the entropy field orthogonal to the classical trajectory. Here $ \theta = \tan^{-1}(\psi'/\phi')$ is the angle between $\phi$ and $\psi$ in the field space  \cite{Gordon:2000hv}.  As can be verified from our numerical analysis we are in the limit where the trajectory in $\psi-\phi$ space is very flat, corresponding to $\theta , \theta' <1$. We will further  elaborate on this point in section \ref{curvature}.
This implies that $\delta s_\K \simeq \delta \psi_{\K}$ and ${\cal S_\K} \equiv \frac{H}{\dot \sigma} \delta s_\K \simeq  \frac{H}{\dot \phi} \delta \psi_\K$ which, using \eqref{approxdeltapsi+}, results in
\ba
\label{ps}
{\cal P_S}  \simeq \left( \dfrac{H}{\dot{\phi}}\right)^2 \dfrac{4\pi \, k^3}{(2 \pi)^3} \, \vert c_2\vert^2 \, Y_\nu^2 \left(\dfrac{\sqrt{\tilde \beta}}{r}e^{-r \, n} \right)  e^{-3n}    \, .
\ea
The $k$-dependence of the entropy power spectrum is only due to the pre-factor $k^3$ and the constant of integration $c_2$. Using \eqref{approxmatch} one has $\mathrm{d} \ln k=\mathrm{d} n_m$ so the spectral index of entropy perturbation, $n_s$, in terms of $n_m$ is
\ba 
n_{s}-1=3+\dfrac{\mathrm{d}\ln \vert c_2 \vert^2}{\mathrm{d} n_m} \, .
\ea
In order to compute the spectral index analytically we use the approximate expression for $c_2$ given by Eq. \eqref{finalapproxc} and  the limiting behavior of Bessel function given by Eq. \eqref{finalapproxbessel} to obtain
\ba
\label{simplifiedns}
n_s-1 \simeq \dfrac{4 \beta_0}{3} \left( \dfrac{\beta_0 }{9}-r n_m\right)  \, .
\ea
We have checked that this gives qualitatively a good approximation for $n_s$ when compared to the full numerical analysis.  As explained before, $n_m$ can be either positive or negative. For  modes which leave the horizon before the phase transition $n_m<0$ whereas for modes leaving the horizon after the phase transition $n_m>0$. This means that $n_s$ can change from a blue spectrum to a red spectrum, depending on whether the mode of interest leaves the horizon before the phase transition or after the phase transition.  This conclusion has been verified numerically.

\section{Back-reactions: Quantum or Classical ?}
\label{back-reactions}

Before we proceed to calculate the curvature perturbation and its power spectrum, we have to determine whether or not the quantum back-reactions are small compared to the classical back-reactions. 
In standard hybrid inflation model with $\beta \gg 1$, corresponding to a very sharp phase transition,  it was shown in \cite{Lyth:2010ch, Abolhasani:2010kr, Fonseca:2010nk, Gong:2010zf} that  the quantum back-reactions from very small scales inhomogeneities produced during the waterfall phase transition dominate exponentially over the classical back-reactions. The back-reactions of quantum fluctuations  uplift the tachyonic instability during the waterfall phase transition  and shuts off inflation very efficiently. 
In our case at hand with $\beta \lesssim 1$, corresponding to a mild phase transition, 
one may expect that depending on model parameters the quantum back-reactions are sub-leading and one can only use the classical back-reactions induced from $\lambda \psi^4$ and $g^2 \psi^2 \phi^2$ interactions  to terminate inflation.  As shown in \cite{Abolhasani:2010kr}, the latter becomes important slightly sooner than the former.

The expectation value of the quantum fluctuations sometime after the phase transition is  
\ba
\label{qm1}
\langle \delta \psi^2 \rangle = \int \dfrac{\mathrm{d}^3 k}{2\pi^3}\, \delta \psi^2_{\mathbf{k}} \, . 
\ea
As mentioned before  $\theta \ll 1$ during most of inflationary period and $\delta s_\K \simeq \delta \psi_\K$. We already calculated $\delta \psi_\K$ for super-horizon modes, given by 
Eq. (\ref{approxdeltapsi+}). We also note that for super-horizon modes $\delta \psi_\K$ evolve
as the background $\psi$ field given by Eq. (\ref{psiapp1}) such that  
$\delta s_\K (N) \equiv \Omega(k) \psi(N)$ where 
\ba
\label{omega}
\Omega(k) \simeq \dfrac{c_2(k)}{\psi_i}~Y_{\nu}\left( \frac{\sqrt{\tilde{\beta}}}{r} \,e^{rN_c}  \right) ~e^{3 N_c/2}
\ea
Here we provide an approximation for $\Omega(k)$ which helps us to evaluate the integral in Eq. (\ref{qm1}).  Using  Eq. \ref{finalapproxc} one has
\ba
\vert c_2 \vert \simeq \gamma(x) \sqrt{\dfrac{2}{k_c}}\,\dfrac{\pi e^{-N_c}}{8rz^{3/2}} ~ J_\nu(x) \, ,
\ea
in which 
\ba
\label{xzdefine}
x \equiv \dfrac{\sqrt{\beta_0 (\Gamma+1)}}{r} e^{- r \, n_m} \gg 1
\quad , \quad
z \equiv \dfrac{k}{k_c} e^{-n_m} \simeq \sqrt{2 \beta_0 r} < 1 \, ,
\ea
and 
\ba
\label{gamma}
\gamma(x)\equiv  \left[r (i+z) \left( \dfrac{2 \nu^2-x^2}{ \nu} \right)+3 z+3 i \right] 
\simeq 6 i + {\cal O}(\beta^2) \, .
\ea
One can check that the main $k$-dependence of $c_2$ comes from the Bessel function and 
in our approximations
\ba
\label{omegasim}
\Omega(k) \simeq \dfrac{H_0}{\sqrt{2k^3}} \frac{6}{4 r \nu \psi_i} 
\simeq \dfrac{H_0}{\sqrt{2k^3}  \, \psi_i} \, .
\ea

As a measure of the strength of the quantum back-reactions, we calculate the ratio 
$\dfrac{\langle \delta \psi^2(n) \rangle}{\psi^2(n)}$ and see under what conditions this ratio is small so one can safely neglect the quantum back-reactions.  Using the above approximations one has
\ba
\label{back0}
\dfrac{\langle \delta \psi^2 \rangle}{\psi^2}\simeq \dfrac{H_0^2}{\psi_i^2} \int_{k_i}^{k_f} \dfrac{\mathrm{d}^3 k}{(2 \pi)^3} \dfrac{1}{2k^3}  = 
\dfrac{H_0^2}{4 \pi^2 \, \psi_i^2} \, \ln{\dfrac{k_f}{k_i}} \, .
\ea
Here  $k_i$ and $k_f$,  respectively, correspond to the largest and smallest modes which become tachyonic during inflation.  For the smallest scale which becomes tachyonic during inflation we have $k_f \simeq \, \sqrt{2\beta r n_f}\,\exp\left(N_f \right) H $ and  for the largest mode we can set $k_i = H$. Plugging these into Eq. (\ref{back0}) and noting that   
$\sqrt{2\beta r n_f} \lesssim 1$ results in  
\ba
\label{back1}
\dfrac{\langle \delta \psi^2 \rangle}{\psi^2}\simeq \dfrac{H_0^2}{\psi_i^2} \dfrac{N_f}{4 \pi^2} \, .
\ea
This equation shows that the quantum back-reactions can be safely ignored if $H_0 < \psi_i$,
that is if one starts with large enough classical waterfall field values at the start of inflation. 
It would be more instructive to express the ratio $H_0^2/\psi_i^2$ in terms of the number of e-foldings and the mass parameters.  Using Eq. (\ref{psiapp1}) one obtains 
\ba
\dfrac{H_0^2}{\psi_i^2} \simeq \dfrac{g^2}{\alpha} e^{-3N_f} ~ \left[ \dfrac{Y_{\nu} \left( \frac{\sqrt{\tilde{\beta}}}{r} e^{-rn_f} \right)}{Y_{\nu} \left( \frac{\sqrt{\tilde{\beta}}}{r} e^{rN_c} \right)} \right]^2 \, .
\ea  
To get this relation, it was assumed that the end of inflation is determined by the back-reactions of the waterfall field on the inflaton field \cite{Abolhasani:2010kr} so $g^2 \psi_f^2 \sim m^2 \phi^2$.  Using the approximations for the Bessel functions given in Eq. (\ref{finalapproxbessel}), the ratio above  is simplified to
\ba
\dfrac{H_0^2}{\psi_i^2} &\simeq& \dfrac{g^2}{\alpha} \exp\left[ (2r\nu -r -3)N_f -  \dfrac{\beta}{2 \nu r^2} e^{2rN_c} \right]  \nonumber\\
&\simeq& \frac{g^2}{\alpha} \exp\left[ \dfrac{2\beta}{3} 
\left( N_f -\dfrac{1}{2r} \right) \right] \, .
\ea
Combined with Eq. (\ref{back1}), the condition under which one can safely neglect the quantum back-reactions till the end of inflation is translated into
\ba
\label{back-final}
\frac{g^2}{\alpha} \ll  \exp\left[ -\dfrac{2\beta}{3} 
\left( N_f -\dfrac{1}{2r} \right) \right] \sim
\exp{ \left[- 30 \beta  \right]    } \, ,
\ea
where the final approximation is for typical values of $r$ and $N_f$ used in our numerical analysis.  For our numerical example with $\beta =0.7$, $\alpha \simeq 0.04$ and $g^2 =2 \times 10^{-12}$ this condition can be met easily.

Eq. (\ref{back-final}) indicates that the strength of the quantum back-reactions is exponentially sensitive to the parameter $\beta$. For fixed values of the inflaton mass and coupling $g$, one has to start with small enough parameter $\beta$ such that the quantum back-reactions can be safely ignored. This   conclusion  is consistent with our starting intuition that  if the phase transition is mild enough  one can neglect the quantum back-reactions. This is also consistent with the conclusion drawn in the model of
standard hybrid inflation with $\beta \gg 1$ that the quantum back-reactions dominate exponentially over the classical back-reactions \cite{Abolhasani:2010kr}.

\section{Curvature Perturbations}
\label{curvature}

We now have all the materials to calculate the final power spectrum of curvature perturbations. For this purpose we need  to know the amplitude of adiabatic curvature perturbations at the time of phase transition as the initial condition and integrate the
evolutions of curvature perturbation from the time of phase transition till end of inflation.
The final amplitude of curvature perturbation, therefore,  is 
\ba
{\cal R}_f = {\cal R}_0 + \int_{0}^{n_f} {\cal R'} \mathrm{d}n \, ,
\ea
where ${\cal R}_0$ represents the adiabatic curvature perturbations in the absence of entropy perturbations. 

As demonstrated in \cite{Gordon:2000hv} the evolution of curvature perturbations for the super-horizon modes, induced by the entropy perturbations, can be written as 
\ba
\label{R-eq}
{\cal R}'= \dfrac{2 \, \theta'}{\sigma'}~ \delta \,s \, .
\ea
As one can see from above equation both $\theta'$ and $\delta s$ can source the curvature perturbations. We also note that $\theta'$ represents the acceleration of $\psi$, specially during the phase transition. As can be seen from our full numerical analysis, the classical background  is such that during inflation and phase transition, $\theta, \theta' \ll 1$.  Inflation ends when the classical back-reactions from   $g^2 \phi^2 \psi^2$ and 
$\lambda \psi^4$ interactions induce large masses for $\phi$ and $\psi$ such that they roll rapidly to the global minimum. Therefore, in the analysis below, we work in the limit where $\theta, \theta' \ll 1$ and consider the end of inflation when $\theta = \theta_f \simeq 1$. 

To calculate the evolution of curvature perturbation from Eq. (\ref{R-eq}) we need to estimate
$\theta'$ and $\delta s_\K$. The derivative of $\theta$ in field space is
\ba
\label{theta-eq}
\theta ' \simeq \tan \theta ' = \dfrac{\psi''}{\phi'} - \dfrac{\psi ' \phi''}{\phi'^2} \, .
\ea
Since $r \simeq 1/N_e \ll 1$, the first term is  much larger than the second term by a factor of $~1/r$ and $\theta' \simeq \psi''/\phi'$.  Furthermore, as mentioned before, 
$\delta s_\K = \Omega (k) \delta \psi_\K $ for $\theta, \theta' \ll 1$ where $\Omega (k)$ is given by Eq. (\ref{omega}).  Combining the above expressions for $\theta'$ and $\delta s$, the final curvature perturbation is given by
\ba
\label{RNe}
{\cal R} (n_f)\simeq {\cal R}_0  -2 \int_{n_{\ast}}^{n_f} \dfrac{\Omega(k)}{\phi'^2} \psi'' \psi ~\mathrm{d}n \, .
\ea
There are some comments in order before we move forward. First, one can check that the 
integrand above scales like  $e^{2 \beta n/3r}$
which is fast growing so one can safely ignore the contribution from the lower limit of the integral. Second, almost all $k$-dependence in the expression above comes from $\Omega(k)$. This means that the main contribution to the spectral index is induced from the entropy perturbations. In other words, the spectral index of the curvature perturbations and the entropy perturbations are more or less the same. 

Now we proceed to approximately evaluate the integral in Eq. (\ref{RNe}). For this purpose note that $\phi'$ scales like $\exp(r n)$ which is nearly constant  so it can be taken out of the integral. Also we can use the background $\psi$ equation,  Eq. (\ref{psi-substituted}), 
to replace $\psi''$ in favors of $\psi$ and $\psi'$. After these simplifications, one obtains
\ba
\label{RNe0}
{\cal R} (n_f)\simeq {\cal R}_0  +\dfrac{2\Omega(k)}{\phi'^2} \left[ \int_{0}^{n_f} 3 \psi \psi' \mathrm{d}n \,+\tilde{ \beta} \left( e^{-2r n_f}-1 \right) \int_{0}^{n_f}  \psi^2 \mathrm{d}n \right] \, ,
\ea
in which the function $\tilde \beta \left( e^{-2rn}-1 \right)$ is taken out of the integral by the same reasoning as for $\phi'$. The first integral is a total derivative which can be calculated easily. To calculate the second integral, note that from Eqs. \ref{psiapp1} and \ref{JY-prime} one has \ba
\label{psidot-psi-app}
\psi \simeq \psi' ~ \left( -3/2 + r \nu - \dfrac{\tilde{\beta} \, e^{-2rn}}{2r \nu}   \right)^{-1}  \, .
\ea
This can be used to transform the second integral above into an approximate total derivative 
containing $\psi \psi'$.  With these simplifications employed, one obtains 
 \ba
\tilde{ \beta} \left( e^{-2r n}-1 \right) \int ^{n} \psi^2 \mathrm{d}n \simeq - \left(3+ \dfrac{\tilde{\beta}}{3} \left(1-e^{-2r n} \right) \right) \dfrac{\psi^2(n)}{2}.
\ea
Plugging this into Eq. (\ref{RNe0}) yields our analytic formula for the curvature perturbation
\ba
\label{RNe2}
{\cal R} (n)= {\cal R}_0  + \dfrac{\tilde{\beta}}{3}\left(1-e^{-2rn}\right)\,\Omega(k)\dfrac{\psi^2}{\phi'^2} \, ,
\ea
where $\Omega(k)$ is given in Eq. (\ref{omega}). 

\begin{figure}
\centerline{\includegraphics[scale=.6]{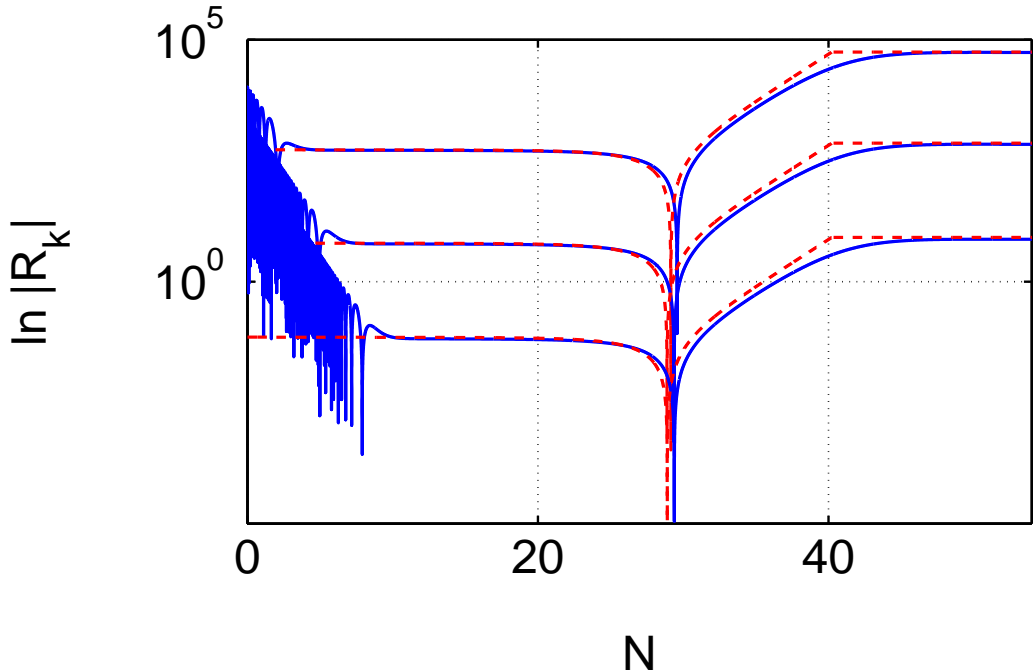}}
\caption{
Here we plot $\ln | R_k|$ for different modes. The  blue solid curves are obtained from the full numerical analysis whereas the red dashed curves are obtained from our analytical formula, Eq. (\ref{RNe2}), setting $\theta_f = .85$. The induced curvature perturbations from the entropy perturbations dominate over the initial adiabatic curvature perturbations
by about two orders of magnitudes.  Note that the apparent singularity at $N\simeq 30$
is due to the fact that $R_k $ vanishes at this point so $\ln | R_k|$ diverges, otherwise it has no physically significant meaning.  From top to bottom, the curves correspond to modes which leave the horizon at $N=3$, $N=6$ and $N=9$ e-foldings respectively. The waterfall phase transition happens at $N_c =7$. 
The parameters are  $M=7.8 \times 10^{-7} \mpl$, $m=2.5 \times 10^{-7} \mpl$ and $g^2 =2 \lambda = 2 \times 10^{-12}$.} 
\vspace{0.7cm}
\label{Rkf} 
\end{figure}

In {\bf Fig.}  \ref{Rkf} we have plotted the predictions of the curvature perturbations from Eq. (\ref{RNe2}) and compared them with the full numerical results. As can be seen, they are in good agreements. Furthermore, the induced curvature perturbations from the entropy perturbations dominates by about two orders of magnitudes over the initial adiabatic perturbations.  Below we find an analytic expression for this enhancement factor.


Now the important question is ``what is the final amplitude of the curvature perturbation?". 
 As long as we are not concerned about the curvature perturbations' $k$-dependence, we can find a simple answer for this question. For this purpose we also need to determine when inflation ends and the curvature perturbations saturate. So far we were working in the limit where the classical back-reactions are subdominant and $\theta \ll 1$. Once the back-reactions become important we expect that $\theta$ to increase significantly. Specifically, from                          Eq. (\ref{theta-eq}) one observes that 
\ba
\theta' \simeq \dfrac{\psi''}{\phi'}~\dfrac{1}{1+\tan^2(\theta)} \, ,
\ea
so once $\theta$ increases significantly $\theta'$ vanishes quickly indicating that both fields 
approaching to their minima. This suggests that the time when the curvature perturbations saturate, which is nearly the time of end of inflation, is when 
\ba
\theta \left( n_f \right) \simeq \dfrac{\psi'\left( n_f \right)}{\phi'\left( n_f \right)} \simeq 1 \, .
\ea
Imposing this criteria in Eq. (\ref{RNe2}) and using the approximations $\phi'(n_f) \simeq e^{-rn_f} \phi'(n_{\ast})$ and $\psi' \simeq \dfrac{\tilde{\beta}}{3} \left( 1-e^{-2rn}\right) \psi$ (derived from  Eq. (\ref{psidot-psi-app}) ) yield the following result for the amplitude of the curvature perturbations at the end of inflation 
\ba
\label{Rf}
{\cal R}_f \simeq {\cal R}_0 \left[ 1+ \dfrac{\psi \left(n_f \right)}{\psi_i} \,e^{rn_f} \right] \, .
\ea 
This is an interesting result. This indicates that the induced curvature perturbations from the entropy perturbations dominates over the initial adiabatic curvature perturbations ${\cal R}_0$
by the factor $ \dfrac{\psi \left(n_f \right)}{\psi_i} \,e^{rn_f} \gg 1$. This enhancement can be seen in {\bf Fig.} \ref{Rkf}. Note however that this enhancing factor can not be arbitrarily large. As can be seen from Eq. (\ref{back1})  the quantum back-reactions can become important  should we start with arbitrarily small $\psi_i$.
It worth mentioning that Eq. (\ref{Rf}) has no precise $k$-dependence and if one is interested in $k$-dependence of the curvature perturbation one should use the original formula 
Eq. (\ref{RNe2}) with the $k$-dependence dictated by $\Omega(k)$.

\begin{figure}
\centerline{\includegraphics[scale=.4]{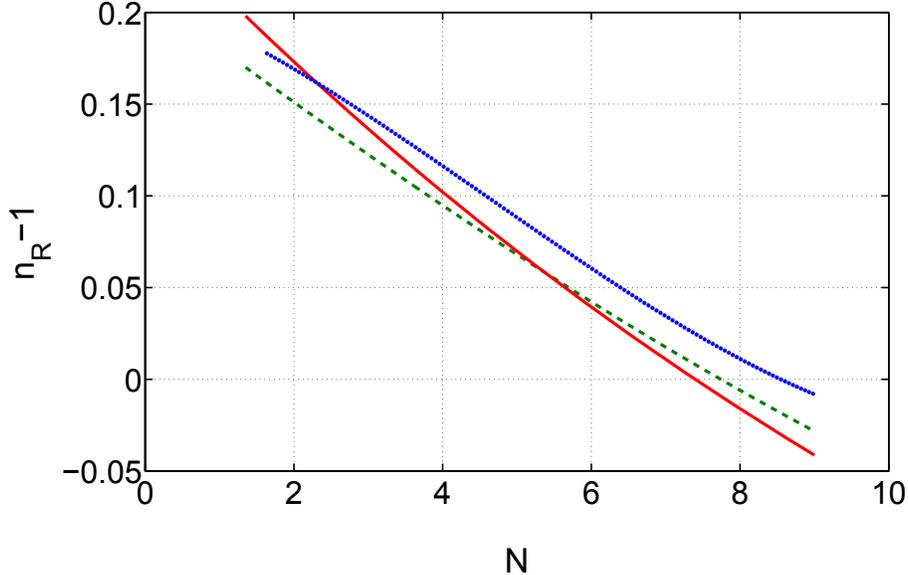}}
\caption{
The spectral index of curvature perturbations, $n_{\cal R} -1$, for modes which leave the horizon at e-folding $N$. The blue dotted curve is obtained from the full numerical analysis whereas the  dashed green curve and the thin solid red curve, respectively, are obtained from our perturbative analysis,  Eq. (\ref{RNe2}), and the $\delta N$ formalism, Eq. (\ref{ns-deltaN}). The parameters are as in {\bf Fig.} \ref{Rkf}. 
As explained in the text, for modes which  leave the horizon approximately before the phase transition ($N_c=7$) one has a blue spectrum whereas for modes which leave the horizon approximately after the phase transition they become red-tilted.  }
\vspace{1cm}
\label{ns} 
\end{figure}

In {\bf Fig.} \ref{ns} we have plotted the spectral index of curvature perturbations, 
$n_{\cal R}-1$, obtained from our analytical formula, Eq. (\ref{RNe2}), compared with the full numerical analysis. As can be seen they are in good agreements. 
Both curves indicate the running from a blue spectrum to a red spectrum, depending on whether the mode of interest leaves the horizon before the phase transition or after the phase transition.  This is in light of discussions below Eq. (\ref{simplifiedns}).

\section{$\delta N$ formalism and   Non-Gaussianity }

\label{deltaN}

In the previous sections we have employed the perturbative approaches in details 
to calculate the curvature perturbations induced from the entropy perturbations. In this section we use the complementary $\delta N$ formalism to verify the previous results for the curvature perturbations. This also enables us to calculate the non-Gaussianity parameter $f_{NL}$ in our model directly. 

In $\delta N$ formalism \cite{Starobinsky:1986fxa, Salopek:1990jq, Sasaki:1995aw, Sasaki:1998ug, Wands:2000dp, Lyth:2004gb, Lyth:2005fi}
curvature perturbation  on super-horizon scales 
can be determined by the variation of number of e-folds with respect to the field values at the time of  horizon crossing
\ba
 \zeta \simeq {\cal{R}} \simeq \delta N   \, ,
\ea 
in which
\ba
\delta N \equiv N \left( \bar\phi + \delta\phi , \bar\psi + \delta\psi  \right)-\bar{N}( \bar\phi, \bar\psi)  \, .
\ea
Here $\bar{N}, \bar \phi$ and $\bar \psi$ respectively are the background number of e-foldings and the background fields starting from the initial flat hyper-surface to the final constant energy density hyper-surface.  One should note that in $\delta N$ formalism the variation should be performed with respect to the field values at the horizon crossing. It is also worth noting that 
in our case the hyper-surface of end of inflation is nearly the same as the surface of constant energy density. This is because the former hyper-surface is given by the relation $\psi^2=\dfrac{\alpha {H}^2}{g^2}$ where the back-reaction of the waterfall field on the inflaton field terminates the slow-roll condition \cite{Abolhasani:2010kr}. From our previous analysis we know that the main part of energy at the end of inflation is due to $\psi$ field so a constant $\psi$ hyper-surface nearly coincides with a constant energy density hyper-surface. 

To employ the $\delta N$ formalism we have to use the background classical trajectory. 
Using Eq. (\eqref{psiapp1}) one can obtain an implicit function of $n$ in terms of fields at the time of horizon crossing
\ba
\psi(n) \simeq \psi_\ast ~ e^{-3 \left(n-n_\ast \right)/2} ~\dfrac{Y_{\nu} 
\left(\frac{\sqrt{\tilde{\beta}} }{r} e^{-rn} \right)}{Y_{\nu} \left( \frac{\sqrt{\tilde{\beta}}}{r} e^{-r n_\ast} \right)} \, ,
\ea
in which $\psi_\ast$ is the value of $\psi$ at horizon crossing, $n_*$. The variation of the above equation results in:
\ba 
\label{n-psi}
n_\psi \equiv \dfrac{\rm{d}n}{\rm{d}\psi_\ast}=\dfrac{1}{\psi_\ast} \dfrac{1}{f(n)} \, ,
\ea
where
\ba
f(n)\equiv\dfrac{3}{2}+\sqrt{\tilde{\beta}} e^{-r n} \dfrac{ Y_{\nu}^{'} \left( \frac{\sqrt{\tilde{\beta}}}{r} e^{-rn} \right)}{Y_{\nu} \left(\frac{\sqrt{\tilde{\beta}}}{r} e^{-r n} \right)}  \, . 
\ea
Here the prime  denotes derivative with respect to the argument of the Bessel function. 
At leading order the curvature perturbation at the end of inflation is given by
\ba 
{\cal{R}} \simeq \frac{\partial n_e}{\partial \psi_*}  \delta \psi_{\ast} \, ,
\ea
where $\delta \psi_{\ast}$ is the value of quantum fluctuations at horizon crossing: $\delta \psi_{\ast} = \left( \dfrac{H}{\sqrt{2 \pi k^3}} \right)_{\ast}$.
The power spectrum of curvature perturbation can be obtained by the above equations
${\cal{P}_{R}}= \frac{k^3}{2 \pi^2} \langle  \delta \psi_{\ast} \delta \psi{\ast}     \rangle$
which results in
\ba
{\cal{P}_{R}}=   \left( \dfrac{H}{2 \pi \psi}\right)_\ast ^2 \dfrac{1}{f(n_e)^2} \, ,
\ea
where $n_e\equiv N_e-N_c$. In our case $n_e$ is about 55 or so. This expression for the curvature perturbation should be compared with Eq. (\ref{RNe2})
for the curvature perturbations obtained from the perturbative approach. We have checked that they are in good agreement with themselves and with the full numerical solution. 
From the above expression one can observe that the leading $k$-dependence of curvature perturbation comes from  $1/\psi_\ast$ factor, so the spectral index of curvature perturbation can be obtained as:
\ba
n_{{\cal R}}-1 \simeq -2 \dfrac{\rm{d} \ln{\psi_\ast}}{\rm{d} ln k}= -2 \dfrac{\rm{d} \ln{\psi_\ast}}{\rm{d}n_\ast} \, .
\ea 
Using Eq. \eqref{n-psi} the final result for spectral index is the following
\ba
\label{ns-deltaN}
n_{{\cal R}} \simeq 1 -2 f(n_\ast) = \dfrac{5}{2}+\sqrt{\tilde{\beta}} e^{-r n_*} \dfrac{Y_{\nu}^{'} \left(\frac{\sqrt{ \tilde{\beta}}}{r} e^{-r n_*} \right)}{Y_{\nu} \left(\frac{\sqrt{\tilde{\beta}}}{r} e^{-r n_*} \right)}  \, . 
\ea
This expression for the spectral index of curvature perturbations should be compared with Eq. (\ref{simplifiedns}), the spectral index of the entropy perturbations obtained from the perturbative approach. 
As demonstrated in previous section, the main source of curvature perturbations comes form the entropy perturbations so the spectral index of curvature perturbation  has the same form as that of the entropy perturbations. Furthermore,  from Eq. (\ref{ns-deltaN}) one observes that for 
modes which leave the horizon before the phase transition, that is $n_* <0$, the spectral index is blue-tilted. However, for modes which leave the horizon (slightly) after the phase transition the spectral index is red-tilted. In {\bf Fig.} \ref{ns} we have plotted the spectral index 
obtained from Eq. (\ref{ns-deltaN}) and compared it with the spectral index obtained from the full numerical analysis and from the perturbative analysis, Eq. (\ref{RNe2}). The running from a blue spectrum to a red spectrum is common in all three curves.

\begin{figure} 
\vspace{-0.5cm}
\centerline{\includegraphics[scale=.6]{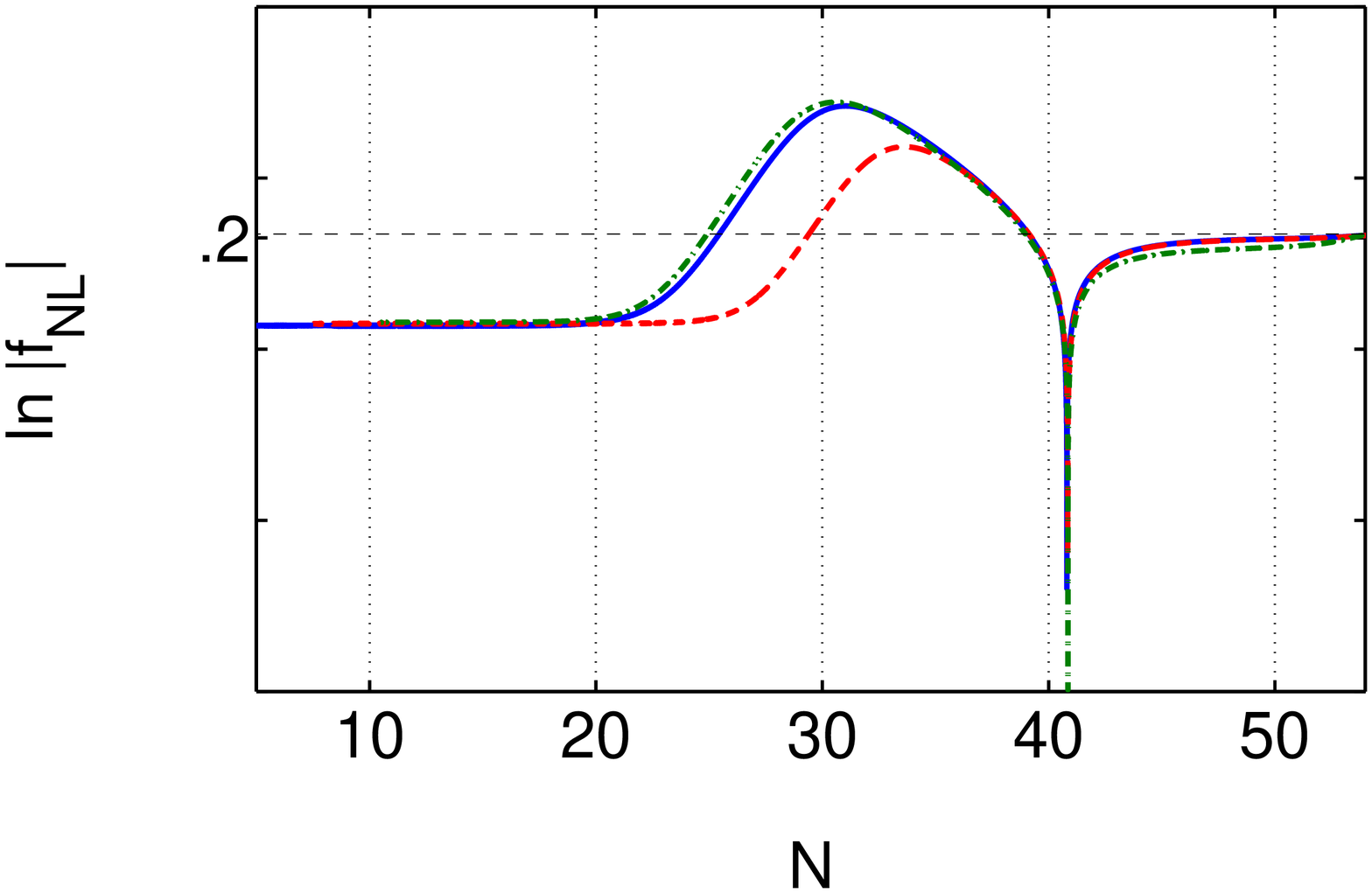}}
\vspace{0cm}
\caption{ Here we present $\ln |f_{NL}|$ for different modes obtained from the full numerical analysis. The solid blue curve, the dashed red curve and the dashed-dotted green curve, respectively, correspond to modes which leave the horizon at $N=3$, $N=6$ and $N=9$
e-foldings.  As can be seen, the final value of $f_{NL}$ saturates at $f_{NL}\simeq 0.2$
which is very well approximated by our analytical formula Eq. (\ref{fNl-analytic}). The bump
at $N\simeq 30$ is because $R_k =0$ at this point and since $f_{NL} \sim R_k^{-1}$  so it peaks at this point. Otherwise, it has no physically significant meaning. Similarly, the apparent
singularity at $N_\simeq 40$ is because $f_{NL}$ changes sign at the final stage of inflation
and $\ln |f_{NL}|$ diverges at this point. As before, it has no significant meaning because we measure  $f_{NL}$ and $R_{k}$ at the end of inflation.
The parameters are the same as in {\bf Fig.} \ref{Rkf}. 
 }
 \vspace{1cm}
\label{fNL} 
\end{figure}


Finally we can easily obtain the non-Gaussianity parameter in our model by the following formula:
\ba
\dfrac{6}{5} f_{NL} = \dfrac{n_{\psi \psi}}{n_{\psi}^2} \simeq - f(n_e) \, .
\ea
Using the approximate behavior of Bessel functions given in the appendix, 
Eqs. \ref{finalapproxbessel} and  (\ref{JY-prime}), to simplify $f(n_e)$ we obtain
\ba
\label{fNl-analytic}
f_{NL}  \simeq \frac{5}{6} \left( r\nu -\dfrac{3}{2} \right) \simeq \frac{5 \beta}{18} \, .
\ea
This is a very interesting formula indicating that the level of non-Gaussianity in our model is controlled by the parameter $\beta$ measuring the sharpness of the phase transition during inflation.  For our numerical example with $\beta =0.7$, we obtain  $f_{NL} \simeq  0.2$. In {\bf Fig.} \ref{fNL} we have plotted $f_{NL}$ obtained from our full numerical analysis for different modes. As we see, $f_{NL}$ measured at the end of inflation is very well approximated by our analytical estimation  Eq. (\ref{fNl-analytic}) and saturates at the expected value $f_{NL} =0.2$. Also as can be seen from the plot, $f_{NL}$ has no significant $k$-dependence. 

We see that generically $f_{NL} \lesssim 1$ in  our model  because of the assumption that the phase transition is mild during inflation. However,  this level of non-Gaussianity  is  much bigger than non-Gaussianity predicted  in simple inflationary models which is at the order of the slow-roll parameters.  Increasing $\beta$ by one or two orders of
magnitude one can obtain significant non-Gaussianities \cite{Gong:2010zf}. However, with $\beta \gg 1$ we approach the standard hybrid inflation model where the waterfall phase transition is very sharp and  the quantum back-reactions dominate exponentially over the classical back-reactions  and one should take their effects into account.  This conclusion is also consistent with the bound obtained in Eq. (\ref{back-final}) in order to neglect the quantum back-reactions compared to the classical back-reactions.


\section{Conclusion and Discussions}
\label{conclusion}
In this work we considered a variant of hybrid inflation where the waterfall phase transition
is mild such that a long period of inflation can be obtained after the phase transition. We found the model parameters where the quantum back-reactions are sub-dominant compared to the classical back-reactions. As can be seen from Eq. (\ref{back-final}) the strength of quantum back-reactions is exponentially sensitive to the parameter $\beta$. For a fixed values of inflaton mass and the coupling $g$, one has to start with small enough $\beta$
such that the quantum-back-reactions are negligible. 

We have calculated the curvature perturbations and the spectral index perturbatively  as well as
 using the complementary $\delta N$ formalism.  We have shown that the curvature perturbations induced from the entropy perturbations dominate over the adiabatic perturbations by  the enhancing factor 
$\frac{\psi_f }{\psi_i} e^{r n_f} \gg 1$.  Consequently,  the spectral index of curvature perturbations has the same form as the spectral index of the entropy perturbations. We have shown that $n_{\cal R}$ runs from a blue spectrum to a red spectrum depending on the time when the mode of interest leaves the horizon. For the modes which leave the horizon approximately  before the phase transition $n_{\cal R} > 1$ whereas for the modes which leave the horizon approximately after the phase transition $n_{\cal R} <1$. This may have interesting consequences when compared to the WMAP data \cite{Komatsu:2010fb} to fit the CMB observations. We would like to come back to this question in future. 

Using the $\delta N$ formalism we have calculated the level of non-Gaussianity in our model. We found the interesting result that $f_{NL} \simeq \frac{5 \beta}{18} \lesssim 1$. This indicates that the parameter $\beta$, which is a measure of the sharpness of the phase transition,  controls not only the strength of the quantum back-reactions but also the level of non-Gaussianity. Although this level of non-Gaussianity may not be detectable observationally
in near future, but it is one order of magnitude larger than the level of non-Gaussianity predicted in simple models of inflation.

As studied in \cite{GarciaBellido:1996qt}, in our model with $\beta \lesssim 1$, primordial black holes can be formed copiously which may be dangerous cosmologically. The analysis of black hole formation and their cosmological consequences are beyond the scope of this work. We would like to come back to this question in future.

\section*{Acknowledgement}

We would like to thank  P. Creminelli, J. Gong,  M. Sasaki, T. Tanaka and  J. Yokoyama for useful discussions and correspondences.
H.F. would like to thank  Yukawa Institute for Theoretical Physics (YITP) for the hospitalities
during  the activities ``Gravity and Cosmology 2010" and ``YKIS2010 Symposium: Cosmology -- The Next Generation"  where this work was in progress. A. A. A. also would like to thank
YITP for the hospitality where this work was in its final stage.

\vspace{1cm}

\appendix
\section{Approximate behaviors of Bessel functions}

In this appendix we present some asymptotic behaviors of Bessel functions for 
the range of parameters relevant to our analysis. These approximate relations are very useful in evaluating  $c_2$ in Eq. \eqref{c} which is also used to calculate the  curvature perturbations power spectrum and the spectral index. 

The Bessel functions $J_\nu(x)$ and $Y_\nu(x)$ satisfy the following equation of motion
\ba
\label{ode}
f''(x)+\dfrac{1}{x} f'(x)+\left( 1-\dfrac{\nu^2}{x^2} \right)f(x)=0 \, .
\ea
We are in the limit where $\nu \gtrsim x \gg 1$. In this limit  the above equation simplifies approximately to
\ba 
f''(x)-\dfrac{\nu^2}{x^2} f(x) \simeq 0 \, ,
\ea
which has a simple solution of
\ba 
\label{firstapproxbessel}
f(x) \simeq c \, x^{ \pm  \left( \sqrt{4 \nu^2 +1}+1 \right)/2} \sim c\, x^{\pm \nu  } \, .
\ea
Here $c$ is a constant of integration and throughout this appendix the upper and lower signs belong to the Bessel function of the first and second kind, respectively. To improve our approximation, note that the main error in the analysis above is in ignoring the factor $1$ in comparison with $\nu^2/x^2$ in \eqref{ode}. Therefore, to next leading order, the differential equation is 
\ba 
\label{approxode}
f''(x)+\left( 1-\dfrac{\nu^2}{x^2} \right)f(x)=0 \, .
\ea
This equation modifies the solution \eqref{firstapproxbessel} by allowing $c$ to be $x$-dependent, whereas previously it was a constant. Since this is the second order approximation we can assume that $c(x)$ is a slowly varying function of $x$ and ignore its second derivative. Using this assumption and substituting \eqref{firstapproxbessel} in to \eqref{approxode} results in the following equation for $c$,
\ba 
\left( \pm \sqrt{4 \nu^2 +1}+1 \right) c'(x) +x \, c(x) \simeq 0 \, ,
\ea
which has a simple solution
\ba 
c(x) \simeq \hat c \exp \left(-\dfrac{x^2}{ \pm 2 \sqrt{4 \nu ^2 +1}+2 } \right)  \simeq \hat 
c \exp \left(- \dfrac{x^2}{4 \nu } \right).  
\ea
Finally, we should determine the constant $\hat c$. Since the overall behavior of our approximate solution looks like the standard small argument limit of 
Bessel function $\left( \nu \sim x \ll 1\right)$, this can be used to fix $\hat c$ so our approximate solution conforms the small argument expansion of the Bessel function. This leads to our desired formulae 
\ba
\label{finalapproxbessel}
J_\nu (x) &\simeq & \dfrac{1}{\Gamma(\nu +1)} \left( \dfrac{x}{2}\right)^{\nu } \exp{\left(-\dfrac{x^2}{4 \nu} \right)}
\nonumber \\
Y_\nu (x) &\simeq & -\dfrac{\Gamma(\nu)}{\pi} \left( \dfrac{x}{2}\right)^{-\nu } \exp{\left(\dfrac{x^2}{4 \nu } \right)} \, .
\ea
One can check numerically that the above equations are good approximation of Bessel functions.

Now we obtain an approximate expression for $c_1$ and $c_2$  in Eq. (\eqref{c}) which we need for the calculation of spectral index and quantum fluctuations. From \eqref{finalapproxbessel} one has,
\ba
\label{JY-prime}
J_\nu ^{'}(x) &\simeq & J_\nu (x) \left(\dfrac{\nu}{ x}-\dfrac{x}{2\nu} \right) 
\nonumber \\
Y_\nu ^{'}(x) &\simeq & Y_\nu (x) \left(-\dfrac{\nu}{ x}+\dfrac{x}{2\nu} \right) \, ,
\ea
where the prime denotes the derivative with respect to the argument $x$. Using these expressions and also the explicit form of Hankel functions one obtains the following approximations for $c_1$ and $c_2$
\ba
\label{finalapproxc} 
c_1 &\simeq & \sqrt{\dfrac{2}{k_c}}~\dfrac{\pi e^{-N_c}}{8r} Y_\nu \dfrac{e^{i z}}{z^{3/2}} \left[-r (i+z) \left( \dfrac{2 \nu^2-x^2}{ \nu} \right) +3 z+3 i \right]
\nonumber \\
c_2 &\simeq & -\sqrt{\dfrac{2}{k_c}}~\dfrac{\pi e^{-N_c}}{8r} J_\nu \dfrac{e^{i z}}{z^{3/2}} \left[r (i+z) \left( \dfrac{2 \nu^2-x^2}{ \nu} \right)+3 z+3 i \right] \, .
\ea
Here $x$ and $z$ are the arguments of Bessel and Hankel functions in Eq. (\eqref{c}), respectively, defined by
\ba
\label{xzdefine}
x &\equiv& \dfrac{\sqrt{\beta_0 (\Gamma+1)}}{r} e^{- r \, n_m} 
\nonumber \\
z &\equiv& \dfrac{k}{k_c} e^{-n_m} \simeq \sqrt{2 \beta_0 r} < 1 \, .
\ea


\end{document}